\documentclass
[prl,superscriptaddress,twocolumn,notitlepage,10pt]{revtex4}%
\usepackage{amsfonts}
\usepackage{amsmath}
\usepackage{amssymb}
\usepackage{graphicx}%
\begin{document}
\textbf{Comment on ``Magnetism of Nanowires Driven by Novel Even-Odd Effects''}

In a recent Letter~\cite{Lounis}, S. Lounis {\it et al.} find that the ground state
of finite antiferromagnetic nanowires deposited on ferromagnets depends on the parity
of the number $N$ of atoms and that a collinear-noncollinear transition exists for odd $N$.
Authors use {\it ab initio} results and a Heisenberg model, which is studied numerically
with an iterative scheme.
In this Comment we argue that the Heisenberg model can much easier
be investigated in terms of a two-dimensional map,
which allows to find an analytic expression for the transition length,
a central result of Ref.~\cite{Lounis} (see their Fig.~3).

\begin{figure}
\includegraphics[width=7cm,clip=yes]{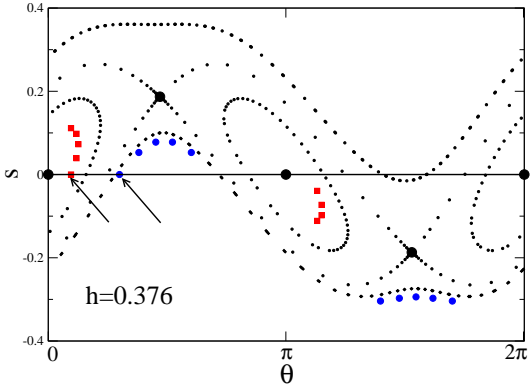}
\caption{
Phase portrait for mapping (1). Red squares and blue circles correspond to configurations for
$N=9$ and $N=10$, respectively. Arrows point to the values $\theta_1$ for the first atom
of the chains.
}
\label{fig1}
\end{figure}

Heisenberg model in Ref.~\cite{Lounis} corresponds to Eq.~(1) of Ref.~\cite{PRL} for $H_A=0$,
which describes antiferromagnetic superlattices in a magnetic field. If we introduce the
variable $s_n=\sin(\theta_n - \theta_{n-1})$, minimization of $H=|J_1|
\sum_{i=1}^{N-1}\cos(\theta_i-\theta_{i+1}) - J_2\sum_{i+1}^N\cos\theta_i$ gives 
\begin{equation}
s_{n+1} = s_n - h\sin\theta_n, \hbox{~~~}
\theta_{n+1} = \theta_n + \sin^{-1}(s_{n+1})
\end{equation}
where $h=J_2/|J_1|$. This is an iterative two-dimensional map whose fixed points of order two
($s_{n+2}=s_n$ and $\theta_{n+2}=\theta_n$) correspond to the ferrimagnetic (FI) configuration 
($(0,0)\leftrightarrow (\pi,0)$) and to the bulk spin-flop state 
($(\bar\theta,\sin 2\bar\theta)\leftrightarrow (-\bar\theta,-\sin 2\bar\theta)$, with
$\cos\bar\theta = h/4$). In Fig.~\ref{fig1} we plot the evolution of the map for different 
initial conditions and $h=0.376$, the special value considered in~\cite{Lounis}. 

Boundary conditions for chains of $N$ atoms are taken into account~\cite{PRL} 
imposing $s_1=0=s_{N+1}$. The determination
of the ground state therefore corresponds to find the value $\theta_1$ such that
iterating the map $N$ times from $(\theta_1,0)$ we get a point on the axis $s=0$. The $N$ values
$\theta_1,\dots,\theta_N$ then give the sought-after configuration.
In Fig.~\ref{fig1} we also plot the first $N$ steps of the map evolution giving the ground states 
for $N=9$ (red squares) and $N=10$ (blue circles), showing 
different behaviors for odd and even $N$. This difference is also visible from
Fig.~6 ($N=52$) and Fig.~10 ($N=53$) in Ref.~\cite{IJMPB}.
Different ground states also reflect on different behaviors for the 
spin wave excitations~\cite{JAP}.

\begin{figure}
\includegraphics[width=7cm,clip=yes]{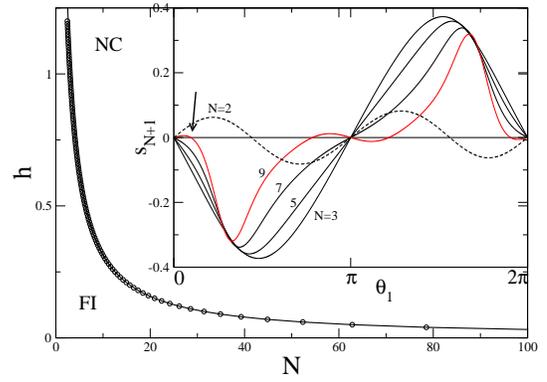}
\caption{
Main: phase diagram for odd-N chains. Inset: we plot $s_{N+1}(\theta_1)$, assuming $s_1=0$, for 
different values of $N$. Arrow points to the value $\theta_1$ for the first atom of the $N=9$
chain.
}
\label{fig2}
\end{figure}

The existence of a minimum length to get a noncollinear configuration for odd $N$
is clear from the inset of Fig.~\ref{fig2}, where we plot $s_{N+1}$ as a function of
$\theta_1$, assuming $s_1=0$. For odd $N<9$, the only zeros are the FI fixed points, but
for $N=9$, $d s_{N+1}/d\theta_1$ changes sign in $\theta_1=0$ and an additional solution
appears: the noncollinear (NC) configuration. For even $N$, non trivial solutions exist
already for $N=2$ (dashed line).

Most importantly, it is possible to linearize the map nearby the fixed points and 
determine the analytical condition for the rising of the NC state. The transition
length is $N_c=\pi/\varphi$, with $\cos\varphi=(2-h^2)/2$ and
$\sin\varphi=h\sqrt{4-h^2}/2$. The curve is plotted with circles in Fig.~\ref{fig2} (main)
along with the asymptotic form $N_c=\pi/h$ (full line) which appears to be a very good
approximation even for small $N$. 

In conclusion, the map method allows to have a direct graphical overview of the
system, to get equilibrium configurations in a fast and reliable way (Fig.~1),
and to find the analytical expression for the transition length (Fig.~2).

Paolo Politi and Maria Gloria Pini

Istituto dei Sistemi Complessi, Consiglio Nazionale delle Ricerche, Via Madonna del Piano 10,
50019 Sesto Fiorentino, Italy

Received \today

\hbox{PACS numbers: 75.75.+a, 75.30.Kz, 75.10.Hk, 05.45.-a}

\end{document}